\documentclass[12pt]{article}
\usepackage{amssymb}
\usepackage{amsfonts}
\usepackage{amsmath}
\usepackage[mathscr]{eucal}
\usepackage{amsthm}
\usepackage{bbold}
\usepackage{bm}
\usepackage{graphicx}
\usepackage{caption}

\theoremstyle{plain}

\numberwithin{equation}{section}
\newcommand{\be}{\begin{equation}}
\newcommand{\ee}{\end{equation}}
\newcommand{\bea}{\begin{eqnarray}}
\newcommand{\eea}{\end{eqnarray}}
\newcommand{\nn}{\nonumber \\}

\newcommand{\p}[1]{(\ref{#1})}

\begin{document}

\begin{titlepage}

\vspace*{0.2cm}

\renewcommand{\thefootnote}{\star}
\begin{center}

{\LARGE\bf Deformed supersymmetric quantum mechanics}\\

\vspace{0.3cm}

{\LARGE\bf with spin variables}\\

\vspace{1.5cm}
\renewcommand{\thefootnote}{$\star$}

{\large\bf Sergey~Fedoruk} \footnote{{\it On leave of absence from V.N.\,Karazin Kharkov National University, Ukraine}},\quad{\large\bf
Evgeny~Ivanov},\quad
{\large\bf Stepan~Sidorov}
 \vspace{0.5cm}

{ \it Bogoliubov Laboratory of Theoretical Physics, JINR,}\\
{\it 141980 Dubna, Moscow region, Russia} \\
\vspace{0.1cm}

{\tt fedoruk@theor.jinr.ru, \; eivanov@theor.jinr.ru, \; sidorovstepan88@gmail.com
}\\
\vspace{0.7cm}

\end{center}
\vspace{0.2cm} \vskip 0.6truecm \nopagebreak

   \begin{abstract}
\noindent
We quantize the one-particle model of the ${\rm SU}(2|1)$ supersymmetric multi-particle mechanics
with the additional semi-dynamical spin degrees of freedom. We find the relevant energy spectrum and the full set of physical states as functions of the mass-dimension
deformation parameter $m$ and ${\rm SU}(2)$ spin $q \in \big( \mathbb{Z}_{>0}$\,, $1/2 + \mathbb{Z}_{\geqslant 0}\big)$\,.
It is found that the states at the fixed energy level form irreducible multiplets of the supergroup ${\rm SU}(2|1)\,$. Also, the hidden
superconformal symmetry ${\rm OSp}(4|2)$ of the model is revealed in the classical and quantum cases.
We calculate the ${\rm OSp}(4|2)$ Casimir operators and demonstrate that the full set of the physical states
belonging to different energy levels at fixed $q$ are unified into an irreducible ${\rm OSp}(4|2)$ multiplet.

\end{abstract}

\vspace{1.5cm}

\phantom{Keywords: }

\newpage

\end{titlepage}

\setcounter{footnote}{0}

\section{Introduction}

In a recent paper \cite{FI16} there was proposed the gauged matrix model of ${\rm SU}(2|1)$ supersymmetric mechanics.
It describes a new ${\cal N}\,{=}\,4$ extension of the Calogero-Moser multi-particle system\footnote{We use the definition ``multi-particle'' instead of ``multi-dimensional'', in order
to avoid a possible confusion with the space-time dimensionality.},
such that the particle mass $m$ is identified with the parameter of deformation of the flat ${\cal N}\,{=}\,4, d\,{=}\,1$ supersymmetry
into curved ${\rm SU}(2|1)$ supersymmetry. Models of ${\rm SU}(2|1)$ supersymmetric mechanics  pioneered in \cite{Sm,BellNer} play the role of
one-dimensional analog of higher-dimensional systems with a rigid curved supersymmetry \cite{FesSei,DumFesSei,SamSor}.
The one-dimensional models possessing such a  deformed supersymmetry and built on different worldline  ${\rm SU}(2|1)$  supermultiplets
were further studied in \cite{IS14a,IS14b,IS15}. All these systems dealt with single ${\rm SU}(2|1)$ multiplets of the fixed type.
There were also studied ${\rm SU}(2|1)$, $d\,{=}\,1$ supersymmetric models \cite{AsDenDz} simultaneously involving chiral $({\bf 2, 4, 2})\,$
and vector $({\bf 3, 4, 1})\,$ supermultiplets. The common feature of this class of the ${\rm SU}(2|1)$ mechanics models is that the Lagrangians for the relevant
physical bosonic $d=1$ fields contain the standard kinetic terms of the second order in the time derivative.

The off-shell construction of \cite{FI16} was based upon a new ${\rm SU}(2|1)$ harmonic superspace with two sets of harmonic variables,
as a generalization of ${\rm SU}(2|1)$ harmonic superspace of ref. \cite{IS15} defined according  to the standard
lines of \cite{HSS0,HSS,IvLech} \footnote{The bi-harmonic formalism of the similar kind was firstly used in  \cite{GIO}
in a different context.}. This bi-harmonic superspace approach provided the proper generalization of the gauging procedure of \cite{DI06,DI07}
and enabled to define ${\rm SU}(2|1)$ analog of the semi-dynamical spin variables of ref. \cite{FIL08}, as well as
to construct the interaction of dynamical and semi-dynamical supermultiplets.

An evident next problem was to quantize the new kind of ${\cal N}\,{=}\,4$ superextended Calogero-Moser systems and inquire whether it inherits
the remarkable quantum properties of its bosonic prototypes.

In the present paper, as the first step towards this goal, we perform quantization of the simplest one-particle case of the multi-particle models constructed in \cite{FI16}.
In this case, the original off-shell formulation involves the superfields describing the ${\rm SU}(2|1)$ multiplets
$({\bf 1, 4, 3})$ and $({\bf 4, 4, 0})\,$, as well as the ``topological'' gauge multiplet which accomplishes
the ${\rm U}(1)$-gauging. After eliminating the auxiliary fields, as the dynamical variables
there remain the $d=1$ bosonic scalar field $x(t)$ and the fermionic ${\rm SU}(2)$-doublet fields $\psi^i(t)$,
$(\psi^i)^\dagger(t)=\bar\psi_i\,$. They come from the multiplet $({\bf 1, 4, 3})\,$.
The bosonic ${\rm SU}(2)$-doublet fields $z^i$, $(z^i)^\dagger=\bar z_i$ coming from the multiplet  $({\bf 4, 4, 0})\,$
describe the semi-dynamical degrees of freedom. The corresponding on-shell component action of the model is written as
\bea
S &=& \int dt\,{\cal L}\,,\nn
{\cal L} &=& \frac12\,{\dot{x}^2}+ \frac{i}{2}\left(\dot{\bar z}_k z^k -
{\bar z}_k \dot z^k\right) - \frac12\,{m^2x^2}-\frac{\left(z^{k}\bar{z}_k\right)^2}{8x^2} + A\left(z^k\bar{z}_k-c\right)\nn
&&+\, \frac{i}{2}\left(\bar\psi_k \dot\psi^k -\dot{\bar\psi}_k \psi^k \right) - m\,\psi^k\bar{\psi}_k
-\frac{\psi^{(i}\bar\psi^{j)} z_{(i} \bar z_{j)}}{x^2} \,.\label{L}
\eea
This action respects local ${\rm U}(1)$ invariance. The ${\rm U}(1)$ gauge field $A$ originates from the topological gauge supermultiplet
and plays the role of Lagrange multiplier for the first class constraint, as we will see in the next Section. Note that only at $c\neq 0$
the Lagrangian \p{L} defines a non-trivial spin-variable generalization of the original model of \cite{Sm,IS14a}: for $c=0$
the constraint effected by $A$ becomes  $z^k\bar{z}_k = 0$ that implies $z^i = 0\,$. At $c\neq 0$ the kinetic term of the bosonic variables $z^i$, $\bar z_i$ in the Lagrangian (1.1) is
of the first-order in the time derivatives, in contrast to the dynamical variable $x$ with the second-order kinetic term. Just for this reason we call $z^i$, $\bar z_i$ semi-dynamical variables
\footnote{As opposed, e.g., to the multiplets exploited in [10], which all are dynamical, with the standard kinetic terms for bosonic and fermionic fields.}. In the Hamiltonian
these variables appear only through the SU(2) current $z_{(i}\bar{z}_{j)}$ and in the interaction terms.

The odd ${\rm SU}(2|1)$ transformations of the component fields leaving  the action \p{L} invariant read
\bea
    &&\delta x =-\,\epsilon_k\psi^k +\bar{\epsilon}^k \bar{\psi}_k\,, \nn
    &&\delta \psi^i = \bar{\epsilon}^i \left(i\dot{x}-m\,x\right)+
\frac{\bar{\epsilon}_k z^{(i}z^{k)}}{x}\,,\qquad
    \delta \bar{\psi}_j =-\,\epsilon_j\left(i\dot{x}+m\,x\right) + \frac{\epsilon^k z_{(j}z_{k)}}{x}
\,,\nn
    &&\delta z^{i} = \frac{z_{k}}{x}\left[\epsilon^{(i}\psi^{k)}+\bar{\epsilon}^{(i}\bar{\psi}^{k)}\right],\qquad
    \delta \bar{z}_{j} =-\,\frac{\bar{z}^{k}}{x}\left[\epsilon_{(j}\psi_{k)}+\bar{\epsilon}_{(j}\bar{\psi}_{k)}\right].\label{tr}
\eea
Their characteristic feature is the presence of the parameter $m$
which can be viewed either as a mass of $d=1$ component fields or as a frequency of the one-dimensional oscillators. It plays the role of the parameter
of deformation of the supersymmetry algebra. While for $m\neq 0$ the transformations \p{tr} generate ${\rm SU}(2|1)$
supersymmetry, in the contraction limit $m=0$ they become those of the flat ${\cal N}=4, d=1$ supersymmetry.

The system \eqref{L} is a  generalization of the ${\cal N}{=}\,4$ superconformally invariant
 models of refs. \cite{FIL08,FIL09,FIL10,FIL12} to the case of non-zero mass. On the other hand,
 in the papers \cite{Pop,HolTop,ISTconf,CHolTop}
(see also \cite{ACCP,BK}) there were found different realizations of ${\cal N}=4, d=1$ superconformal groups and established some relations
between the deformed supersymmetries and superconformal symmetries. One of the purposes  of our paper
is to clarify the role of the deformed ${\rm SU}(2|1)$ supersymmetry and its superconformal extension
in the quantum spectrum of the systems with semi-dynamical variables. In particular, we find out the presence of
hidden superconformal symmetry in the model \eqref{L}, on both the classical and the quantum levels.

We start in Sect.\,2 with presenting the Hamiltonian formulation of the system  \eqref{L}. We find the Noether charges
of the ${\rm SU}(2|1)$ transformations of the component fields. In Sect.\,3, the quantization of the model is performed.
The energy spectrum and stationary wave functions are defined. The latter are characterized by an external ${\rm SU}(2)$ spin $q \in \big(\mathbb{Z}_{>0}$\,,
$1/2 + \mathbb{Z}_{\geqslant 0}\big)$\,,
and involve a non-trivial holomorphic dependence on the semi-dynamical spin variables. All energies also depend on $q\,$.
The states on each energy level belong to an irreducible representation of the supergroup ${\rm SU}(2|1)\,$ (and of its central extension $\widehat{\rm SU}(2|1)\,$,  with
the canonical Hamiltonian as the ``central charge'').
The sets of these representations are defined from the dependence of the wave function on the semi-dynamical variables.
Sect.\,4 is devoted to the analysis of hidden ${\cal N}{=}\,4, d=1$ superconformal symmetry of the model under consideration.
We identify it as the ${\rm OSp}(4|2)$ superconformal symmetry in the ``trigonometric'' realization and present the explicit form
of  its generators. We also compute the quantum ${\rm OSp}(4|2)$ Casimir operator and show that the physical states
with different energy levels and fixed spin $q$ are combined into an irreducible ${\rm OSp}(4|2)$ supermultiplet.  Sect.\,5 collects
the concluding remarks and some problems for the further study.

\section{Hamiltonian analysis and Noether charges}
Performing the Legendre transformation for the Lagrangian \eqref{L}, we obtain the canonical Hamiltonian
\bea
H_C&=&H-A\left( T-c\right),\nn
H &=& \frac12\left( p^2 + m^2 x^2\right) + m\,\psi^k\bar{\psi}_k  -\frac{S^{(ij)}S_{(ij)}}{4x^2}
+ \frac{\psi^{(i}\bar\psi^{j)} S_{(ij)}}{x^2}\,,\label{H}
\eea
where $S^{(ij)}$ are
bilinear combinations of $z^i$ and $\bar{z}_j$ with external
${\rm SU}(2)$ indices,
\begin{equation}\label{S}
S^{(ij)}:= z^{(i}\bar{z}^{j)}\,,
\end{equation}
while $T$ is the ${\rm SU}(2)$-scalar
\begin{equation}\label{fcc}
T := z^{k}\bar{z}_k \,.
\end{equation}

The action (\ref{L}) produces also the primary constraints
\be\label{const-z}
p_{\,z}{}_k+\frac{i}{2}\,\bar z_k\approx 0\,,\qquad
p_{\,\bar z}{}^k-\frac{i}{2}\,  z^k\approx 0\,,
\ee
\be\label{const-psi}
p_{\,\psi}{}_{k}-\frac{i}{2}\,\bar\psi_{k}\approx 0\,,\qquad
p_{\,\bar\psi}{}^k-\frac{i}{2}\,  \psi^k\approx 0\,,
\ee
\be\label{const-A}
p_{A}{} \approx 0\,.
\ee

The constraints (\ref{const-z}), (\ref{const-psi}) are second class and
we are led to introduce Dirac brackets for them. As a result, we eliminate the momenta $p_{\,z}{}_k$, $p_{\,\psi}{}_{k}$ and c.c..
The residual variables possess the following Dirac brackets
\be\label{brackets}
\left\{x, p \right\}^*=1\,,\qquad \left\{z^i, \bar z_j \right\}^* = i\delta_{j}^i\,,\qquad
\left\{\psi^i, \bar{\psi}_{j}\right\}^* = -\,i\delta_{j}^i\,.
\ee
So, the bosonic variables $z^i$, $\bar{z}_j$ describing spin degrees of freedom are self-conjugated phase variables.
The triplet of the quantities \eqref{S} form $su(2)$ algebra with respect to the brackets  \eqref{brackets}:
\be
\left\{S^{(ij)}, S^{(kl)} \right\}^*  = -\,i\left[\epsilon^{ik}S^{(jl)}+\epsilon^{jl}S^{(ik)} \right] .
\ee

Requiring the constraint (\ref{const-A}) to  be  preserved by the Hamiltonian (\ref{H}) results in a secondary constraint
\begin{equation}\label{f-c}
T -c= z^{k}\bar{z}_k-c \approx 0 \,.
\end{equation}
The field $A$ is the Lagrange multiplier for this constraint, as is seen from the
Lagrangian (\ref{L}) and the Hamiltonian (\ref{H}). So in what follows we can omit  the constraint (\ref{const-A}) and
exclude the variables $A$, $p_A$ from the set of phase variables.

Applying Noether procedure to the transformations \eqref{tr} we obtain the following expressions for the ${\rm SU}(2|1)$ supercharges:
\be\label{Q}
    Q^i = \left(p-imx\right)\psi^i + \frac{i}{x}\,S^{(ik)}\psi_{k}\,,\qquad
    \bar{Q}_j = \left(p+imx\right)\bar{\psi}_j-\frac{i}{x}\,S_{(jk)}\bar{\psi}^{k}.
\ee
The supercharges \eqref{Q} generate the centrally-extended superalgebra $\hat{su}(2|1)$ with respect to the Dirac brackets \eqref{brackets}:
\bea
    &&\left\{Q^i, \bar{Q}_j \right\}^*=-\,2i\delta^i_j H-2im\left(I^i_j-\delta^i_j F \right),\nn
    &&\left\{F, Q^i \right\}^*=-\frac{i}{2}\,Q^i,\qquad
    \left\{F, \bar{Q}_j \right\}^*=\frac{i}{2}\,\bar{Q}_j\,,\nn
    &&\left\{I^i_j, Q^k \right\}^*=-\frac{i}{2}\left(\delta_j^k Q^i+\epsilon^{ik}Q_j\right),
\qquad \left\{I^i_j, \bar{Q}_k \right\}^*=
\frac{i}{2}\left(\delta_k^i\bar{Q}_j+\epsilon_{jk}\bar{Q}^i\right),\nn
    &&\left\{I^i_k, I^j_l \right\}^*=i\left(\delta^i_l I^j_k - \delta^j_k I^i_l\right).\label{algebra}
\eea
Here, the $SU(2)$ and $U(1)$ generators are defined as
\be
    I^i_k=\epsilon_{kj}\left[S^{(ij)}+\psi^{(i}\bar{\psi}^{j)}\right]\,,\qquad
    F=\frac12\,\psi^{k}\bar{\psi}_{k}\,.\label{b-gen}
\ee
The Hamiltonian $H$ commutes with all other generators and so can be treated as a central charge operator. As was shown in [8], $\hat{su}(2|1)$ can be represented
as a semi-direct sum $\hat{su}(2|1)\simeq su(2|1) +\!\!\!\!\!\!\supset u(1)$ of the standard $su(2|1)$ superalgebra and an extra R-symmetry generator $F\,$.

Let us compare the model defined above with the model of refs. \cite{IS14a,IS14b}.
As opposed to these papers, here we consider the system with the
spin degrees of freedom added. As a result, the supercharges
\eqref{Q} acquire additional terms with the $su(2)$ generators
$S^{(ij)}\,$, which produce extra conformal potential-like terms in the
Hamiltonian \eqref{H}. Moreover, the generators $S^{(ij)}$ make an
additional contribution to the $su(2)$ generators $I^{(ij)}$ of  the
$\hat{su}(2|1)$ superalgebra \eqref{algebra}.

\section{Quantization}

At the quantum level, the system \eqref{L} is described by the operators
${\mathbf x}$, ${\mathbf p}$; ${\bm{\psi}}^i$, $\bar{\bm{\psi}}_i$;
${\mathbf z}^i$, $\bar{\mathbf z}_i\,$ which obey the algebra
\be\label{alg-1}
\left[{\mathbf x}, {\mathbf p} \right] =i\,,\qquad
\left[{\mathbf z}^k, \bar{\mathbf z}_j \right]= -\delta_{j}^k \,,\qquad
\left\{ {\bm{\psi}}^k, \bar{\bm{\psi}}_j \right\} =\delta_{j}^k \,.
\ee
The quantum counterpart of the first class constraint \eqref{f-c} is the condition that, on the physical wave functions,
\be\label{T-constr-1}
{\mathbf T}-2q \equiv {\mathbf z}^k \bar{\mathbf z}_k -2q\approx 0\,.
\ee
The constant $2q$ in \eqref{T-constr-1} is a counterpart of the classical constant $c$ in \eqref{f-c}; the difference between $c$ and $2q$
can be attributed to the operator ordering ambiguity.

The quantum supercharges generalizing the classical ones \eqref{Q} are uniquely determined as
\be\label{Q-quant}
{\mathbf{Q}}^k = \left({\mathbf p}-im\,{\mathbf x} \right)\,{\bm{\psi}}^k +
\frac{i\,{\mathbf S}^{(kj)}\,{\bm{\psi}}_{j}}{{\mathbf x}}\,,\qquad
\bar{\mathbf{Q}}_k=  \left({\mathbf p}+im\,{\mathbf x} \right)\,\bar{\bm{\psi}}_k -
\frac{i\,{\mathbf S}_{(kj)}\,\bar{\bm{\psi}}^{j}}{{\mathbf x}}\,.
\ee
Their closure is a quantum counterpart of the classical superalgebra \eqref{algebra}:
\bea
    &&\left\{{\bf Q}^i, \bar{\bf Q}_k \right\}=2\delta^i_k {\bf H} +
2m\left({\bf I}^i_k-\delta^i_k {\bf F} \right),\nn
    &&\left[{\bf F}, {\bf Q}^k \right]=\frac{1}{2}\,{\bf Q}^k,\qquad
\left[{\bf F}, \bar{\bf Q}_k \right]=-\frac{1}{2}\,\bar{{\bf Q}}_k\,,\nn
    &&\left[{\bf I}^i_k, {\bf Q}^j \right]=\frac{1}{2}\left(\delta_k^j{\bf Q}^i+\epsilon^{ij}{\bf Q}_k\right),
\qquad \left[{\bf I}^i_k, \bar{\mathcal{{\bf Q}}}_j \right]=
-\frac{1}{2}\left(\delta_j^i\bar{\mathcal{{\bf Q}}}_k+\epsilon_{kj}\bar{\mathcal{{\bf Q}}}^i\right) ,\nn
    &&\left[{\bf I}^i_k, {\bf I}^j_l \right]=\delta^j_k {\bf I}^i_l - \delta^i_l {\bf I}^j_k\,.\label{q-algebra}
\eea
The quantum Weyl-ordered Hamiltonian appearing in \eqref{q-algebra} and generalizing \eqref{H} reads
\begin{eqnarray}
\label{q-H}
{\mathbf H} = \frac12 \left( {\mathbf p}^2 + m^2 {\mathbf x}^2 \right)
-\frac{{\mathbf S}^{(ik)}{\mathbf S}_{(ik)}}{4{\mathbf x}^2} + \frac{m}{2} \left[{\bm{\psi}}^k,\bar{\bm{\psi}}_k\right]
+ \frac{{\bm{\psi}}^{(i}\bar{\bm{\psi}}^{k)} {\mathbf S}_{(ik)}}{{\mathbf x}^2}  \,.
\end{eqnarray}
Here, ${\mathbf S}^{(ik)}={\mathbf z}^{(i} \bar{\mathbf z}^{k)}=\bar{\mathbf z}^{(i} {\mathbf z}^{k)}$
satisfies $\left[{\mathbf S}^{(ij)}, {\mathbf S}^{(kl)} \right]=\epsilon^{ik}{\mathbf S}^{(jl)}+\epsilon^{jl}{\mathbf S}^{(ik)}$.
The remaining even generators in  \eqref{q-algebra} are defined by
\be\label{q-gen}
{\mathbf I}^i_k=\epsilon_{kj}\left({\mathbf S}^{(ij)}+ {\bm{\psi}}^{(i}\bar{\bm{\psi}}^{j)}\right),
\qquad {\mathbf F}=\frac{1}{4}\left[{\bm{\psi}}^k,\bar{\bm{\psi}}_k\right].
\ee

Taking into account that
\be \label{iq-ST}
-\frac12\, {\mathbf S}^{(ik)}{\mathbf S}_{(ik)}=\frac12\,{\mathbf T}\left(\frac12\,{\mathbf T}+1 \right),
\qquad {\mathbf T}\equiv {\mathbf z}^k \bar{\mathbf z}_k\,,
\ee
and also the first class constraint (\ref{T-constr-1}), we observe that
the bosonic Hamiltonian, when applied to the wave functions, can be cast in the form
\be \label{q-Hb}
{\mathbf H}^{bose} = \frac12 \left[ {\mathbf p}^2 + m^2 {\mathbf x}^2
+\frac{q\left(q+1\right)}{{\mathbf x}^2}\right].
\ee
It involves the potential which is a sum of the oscillator potential and
the inverse-square ($d\,{=}\,1$ conformal) one.

\subsection{Casimir operators}
The second- and third-order Casimir operators of $su(2|1)$ are constructed as \cite{IS15}
\bea
    \label{C2}
    {\bf C}_2&=& \left(\frac{1}{m}\,{\bf H} - {\bf F}\right)^2 -\frac12\, {\bf I}^i_k {\bf I}^k_i +\frac{1}{4m}\,\left[{\bf Q}^i,\bar {\bf Q}_i \right],\\
    \label{C3}
    {\bf C}_3&=& \left({\bf C}_2+\frac{1}{2}\right)\left(\frac{1}{m}\,{\bf H} - {\bf F}\right) + \frac{1}{8m}\left\lbrace\delta^j_i\left(\frac{1}{m}\,{\bf H} - {\bf F}\right)-{\bf I}^j_i\right\rbrace\left[{\bf Q}^i,\bar {\bf Q}_j\right].
\end{eqnarray}
Here the combination ${\displaystyle\frac{1}{m}\,\mathbf{H}-\mathbf{F}}$ plays the role of the full internal ${\rm U}(1)$ generator. Both $su(2|1)$ Casimirs involve just this combination, not $\mathbf{H}$
and $\mathbf{F}$ separately.
For the specific realization of $\hat{su}(2|1)$ generators given by \eqref{Q-quant}, \eqref{q-H} and \eqref{q-gen}
the Casimir operators \eqref{C2} and \eqref{C3} take the form
\bea
    \label{su21-Cas2-1}
    {\mathbf C}_2&=& \frac{1}{m^2}\left({\mathbf H}+\frac{m}{2}\right)\left({\mathbf H}-\frac{m}{2}\right) + \frac{1}{2}\,{\mathbf S}^{(ik)}{\mathbf S}_{(ik)}\nn
&=& \frac{1}{m^2}\left({\mathbf H}+\frac{m}{2}\right)\left({\mathbf H}-\frac{m}{2}\right) - q\left(q+1\right),\\
    \label{su21-Cas3-1}
    {\mathbf C}_3&=& \frac{1}{m}\,{\bf H} {\mathbf C}_2\,.
\eea
Their eigenvalues on a wave function characterize the irreducible representation of the supergroup ${\rm SU}(2|1)\,$ to which this wave function belongs.
So we come to the important conclusion that states on a fixed energy level form irreducible multiplets of ${\rm SU}(2|1)$. Such  ${\rm SU}(2|1)$ multiplets can equally
be treated as irreducible multiplets of the extended supergroup $\widehat{{\rm SU}}(2|1)\,$, with ${\mathbf H}$ as an additional Casimir taking the same value
on all states of the given ${\rm SU}(2|1)$ multiplet.

\subsection{Wave functions and energy spectrum}
To define the wave functions, we use the following realization of the basic operators  \eqref{alg-1}:
\be\label{real-alg-1}
{\mathbf x} = x\,,\quad {\mathbf p}= -i\frac{\partial}{\partial x} \,,\qquad
{\mathbf z}^k = z^k\,,\quad \bar{\mathbf z}_k= \frac{\partial}{\partial z^k} \,,\qquad
{\bm{\psi}}^k={\psi}^k \,,\quad \bar{\bm{\psi}}_k = \frac{\partial}{\partial {\psi}^k} \,.
\ee
Here $x$ and $z^k$ are real and complex bosonic commuting variables, respectively,
 whereas ${\psi}^k$ are complex fermionic ones. Then the wave function $\Phi$ can be defined as a superfield depending on $x$, $z^k$ and ${\psi}^k$,
{\it i.e.}, as $\Phi(x,z^k,{\psi}^k)\,$.

The quantum constraint (\ref{T-constr-1}) amounts to the following condition on the physical wave functions
\be\label{T-constr-wf}
z^k\frac{\partial}{\partial z^k}\,\Phi^{(2q)}=2q\,\Phi^{(2q)},
\ee
whence the superscript $(2q)$ for the wave functions of physical states.

The solution of the constraint (\ref{T-constr-wf}) is a monomial of the degree $2q$  with respect to $z^k$.
Thus the constant $q$ is quantized as:
\be\label{q-Z}
2q\in \mathbb{Z}_{\geqslant 0}\,.
\ee
On the other hand, this constant defines the conformal potential [see (\ref{q-Hb})]. Hence,
the strength of the   conformal potential is quantized in our model. The constant $q$ has the meaning of the external
${\rm SU}(2)$ spin. In what follows we will restrict our study to the option $q\,{>}\,0$\, only, since quantization in the simplest case $q\,{=}\,0$
was earlier considered in \cite{Sm} and, in more detail,
in \cite{IS14a}\footnote{The quantum Hamiltonian given in ref. \cite{IS14a} coincides with the quantum Hamiltonian \eqref{q-H}
for $q=0$ up to a constant shift by $m/2$\,.}. No tracks of the spin variables remain at $q\,{=}\,0\,$.

Let us find the energy spectrum of the model, that is, the allowed energy levels $E_{\ell}$ and
the corresponding wave functions $\Phi^{(2q,\ell)}$ determined by the eigenvalue problem
\be\label{eq-wf}
{\mathbf H}\,\Phi^{(2q,\ell)}=E_{\ell}\,\Phi^{(2q,\ell)}\,.
\ee
To this end, we start with the general expression
of the wave function $\Phi^{(2q,\ell)}$ for $q\,{>}\,0$
\begin{eqnarray}\label{wf-1}
    \Phi^{(2q,\ell)}&=&A_+^{(2q,\ell)}+\psi^i B_i^{(2q,\ell)}+\psi^i C_i^{(2q,\ell)}+\psi^i\psi_i\, A_-^{(2q,\ell)},\\
    && A_{\pm}^{(2q,\ell)}= A^{(\ell)}_{\pm(k_1\ldots k_{2q})}z^{k_1}\ldots z^{k_{2q}} , \nonumber \\
    && B_i^{(2q,\ell)} = B^{(\ell)}_{(k_1\ldots k_{2q-1})}\,z_i z^{k_1}\ldots z^{k_{2q-1}},\nn
    && C_i^{(2q,\ell)} = C^{(\ell)}_{(ik_1\ldots k_{2q})}z^{k_1}\ldots z^{k_{2q}}\,.\nonumber
\end{eqnarray}
Here, $A_+^{(2q,\ell)}$, $A_-^{(2q,\ell)}$ stand for the bosonic states with the ${\rm SU}(2)$ spin $q$ and
$B_i^{(2q,\ell)}$, $C_i^{(2q,\ell)}$ for fermionic states with the ${\rm SU}(2)$ spins $q\,{\mp}\,1/2\,$, respectively.

For the component wave functions in (\ref{wf-1}), the eigenvalue problem \eqref{eq-wf} amounts to the following set of equations
\begin{eqnarray}\label{eq-wf-1}
&&  \frac12 \left[ {\mathbf p}^2 + m^2 {\mathbf x}^2
+\frac{q\left(q+1 \right)}{{\mathbf x}^2}\right] A^{(\ell)}_{\pm(k_1\ldots k_{2q})}=\left(E_{\ell}\pm m \right) A^{(\ell)}_{\pm(k_1\ldots k_{2q})}\,,\\
&&  \frac12 \left[{\mathbf p}^2 + m^2 {\mathbf x}^2
+\frac{\left(q+1 \right)\left(q+2\right)}{{\mathbf x}^2}\right] B^{(\ell)}_{(k_1\ldots k_{2q-1})}=
E_{\ell}\,B^{(\ell)}_{(k_1\ldots k_{2q-1})}\,, \label{eq-wf-2} \\
&&  \frac12 \left[ {\mathbf p}^2 + m^2 {\mathbf x}^2
+\frac{q\left(q-1 \right)}{{\mathbf x}^2}\right]C^{(\ell)}_{(ik_1\ldots k_{2q})}=
E_{\ell}\, C^{(\ell)}_{(ik_1\ldots k_{2q})}\,. \label{eq-wf-3}
\end{eqnarray}

Now we should take into account (see, e.g., \cite{Calogero,Per}) that the solution of the stationary Schr\"{o}dinger equation,
\be\label{eq-conf}
\frac12 \left[{\mathbf p}^2 + m^2 {\mathbf x}^2
+\frac{\gamma\left(\gamma-1 \right)}{{\mathbf x}^2}\right] |\ell, \gamma\,\rangle=
\mathscr{E}_\ell\, |\ell, \gamma\,\rangle\,,
\ee
where $\gamma$ is a constant,  is given by
\bea
&& |\ell , \gamma\,\rangle=
x^\gamma\,L^{(\gamma - 1/2)}_\ell(mx^2)\,\exp(-mx^2/2)\,,\qquad \ell=0,1,2,\ldots\,,\label{eq-conf-sol} \\
&&\mathscr{E}_\ell=
m\left(2\ell+\gamma+\frac12 \right), \label{eq-conf-ener}
\eea
where $L^{(\gamma - 1/2)}_\ell$ is a generalized Laguerre polynomial.
The parameter $\gamma$ takes the values $q+1$, $q+2$ and $q$ for the equations \eqref{eq-wf-1}, \eqref{eq-wf-2} and \eqref{eq-wf-3}, respectively.

It is worth noting that there exists another solution of \eqref{eq-conf} given by
\bea
&& \widetilde{|\ell , \gamma\,\rangle} = x^{1-\gamma}\,L^{(1/2 -\gamma)}_\ell(mx^2)\,\exp(-mx^2/2)\,,\qquad \ell=0,1,2,\ldots\,,\label{2sol}\\
&& \mathscr{E}^*_\ell = m\left(2\ell-\gamma+\frac{3}{2}\right).
\eea
It is singular at $x\,{=}\,0$ for $\gamma\,{>}\,1$\,. For  $\gamma\,{=}\,1$ and $\gamma\,{=}\,0$\,, this solution spans, together with \eqref{eq-conf-sol},
the total set of solutions of the harmonic oscillator \cite{Calogero,Per}.  For example, at $\gamma\,{=}\,0$ the solution \eqref{eq-conf-sol}
yields the harmonic oscillator spectrum for even levels $\ell$,  while the second solution \eqref{2sol} corresponds to odd levels. The choice of $\gamma\,{=}\,1$ gives rise to
the equivalent spectrum. When $\gamma\,{=}\,1/2$\,, the solutions \eqref{2sol} and \eqref{eq-conf-sol} coincide.
In order to avoid singularities at $x\,{=}\,0$ for the case of $q\,{>}\,0$  we are interested in here (and thereby ensure the relevant wave functions to be normalizable),
we are led to throw away the solution \eqref{2sol}. Although for $q\,{=}\,1$ this type of solution of eq. \eqref{eq-wf-3} contains no singularity
at $x\,{=}\,0$, the result of action of the supercharges on the relevant state still produces singular solutions.
So, the solution \eqref{2sol} suits only for the case of $q\,{=}\,0$\,, which corresponds to supersymmetric harmonic oscillator without spin variables \cite{Sm,IS14a}.
Note that for $q\,{=}\,0$\,, the equation \eqref{eq-wf-2} is absent, since $B^{(\ell)}\equiv 0$ in this case.

Applying \eqref{eq-conf-sol} to eqs. \eqref{eq-wf-1} - \eqref{eq-wf-3}, we solve them as
\bea
    &&A_{+(k_1\ldots k_{2q})}^{(\ell)}=A_{+(k_1\ldots k_{2q})}\,|\ell , q+1\,\rangle\,,\qquad
    C_{(ik_1\ldots k_{2q})}^{(\ell)}=C_{(ik_1\ldots k_{2q})}\,|\ell , q\,\rangle\,, \label{26u} \\
    &&A_{-(k_1\ldots k_{2q})}^{(\ell)}=A_{-(k_1\ldots k_{2q})}\,|\ell -1, q+1\,\rangle\,,\qquad
    B_{(k_1\ldots k_{2q-1})}^{(\ell)}=B_{(k_1\ldots k_{2q-1})}\,|\ell -1 , q+2\,\rangle\,,\label{27u}
\eea
and determine the relevant energy spectrum
\be\label{eq-conf-ener-comp}
    E_{\ell}= m\left(2\ell + q + \frac12\right),\qquad \ell=0,1,2,\ldots
\ee
Note that the solutions \p{26u} for $A_+^{(2q,\ell)}$ and $C_i^{(2q,\ell)}$ exist for $\ell\,{\geqslant}\,0$\,, while the solutions \p{27u}
for $A_-^{(2q,\ell)}$ and $B_i^{(2q,\ell)}$ exist for $\ell\,{\geqslant}\,1$ only.

We see that the energy spectrum is  discrete. Thus the total wave function at fixed spin $q$ is a superposition
of the wave functions at all energy levels:
\be\label{eq-wf-f}
\Phi^{(2q)}=\sum_{\ell = 0}^{\infty}\Phi^{(2q,\ell)}\,.
\ee
According to eq. \eqref{eq-conf-ener-comp},
at the fixed parameter $q$ to each value of $\ell$ there corresponds the definite energy $E_{\ell}\,$.
Taking into account this property, the expressions \eqref{su21-Cas2-1} and \eqref{su21-Cas3-1} for ${\rm SU}(2|1)$ Casimirs, and also the expression \p{q-gen} for the generator $\mathbf{F}$,
we observe that the states $\Phi^{(2q,\ell)}$ form irreducible multiplets
of the supergroup ${\rm SU}(2|1)$, such that the internal ${\rm U}(1)$ generator ${\displaystyle\frac{1}{m}\,\mathbf{H}-\mathbf{F}}$ has a non-trivial action on the states within each fixed multiplet.
The central charge $\mathbf{H}$ takes the same value on all states of the given ${\rm SU}(2|1)$ multiplet and distinguishes different such  multiplets by ascribing to them different level numbers $\ell$\,.  In this way,
the space of all quantum states splits into  irreducible multiplets of the extended supergroup $\widehat{\rm SU}(2|1)$ (see also the remark in the end of Sect. 3.1).

As the last topic of this Section, we will analyze the precise ${\rm SU}(2|1)$ representation contents of the wave functions for $q\,{>}\,0$\,.
A brief description of ${\rm SU}(2|1)$ representations \cite{repres} can be found in Appendix {\bf C} of \cite{IS15}.

The ground state, $\ell\,{=}\,0$\,, corresponds to the lowest energy
\be
    E_0=m\left(q+\frac12\right).\label{V-energy}
\ee
This energy value is non-vanishing for $q\,{>}\,0$\,. The relevant eigenvalues of the Casimir operators defined in \eqref{su21-Cas2-1} and \eqref{su21-Cas3-1}
can be shown to vanish, {\it i.e.},
\bea
    {\mathbf C}_2\,\Phi^{(2q,0)}=0\,,\qquad {\mathbf C}_3\,\Phi^{(2q,0)}=0\,.
\eea
Thus the ground state corresponds to an atypical  ${\rm SU}(2|1)$ representation containing unequal numbers
of bosonic and fermionic states, with the total dimension $4q\,{+}\,3$\,.
Its wave function $\Omega^{(2q,0)}$ involves, in its component expansion, only two terms
\begin{eqnarray}
\Phi^{(2q,0)} = A_+^{(2q,0)}+\psi^i C_i^{(2q,0)}.\label{Vacuum}
\end{eqnarray}
Indeed, this expansion encompasses $2q\,{+}\,1$ bosonic and $2q\,{+}\,2$ fermionic states. These states can not be simultaneously
annihilated  by the supercharges \eqref{q-gen}.
Hence, ${\rm SU}(2|1)$ supersymmetry is spontaneously broken for $q\,{>}\,0$\,, as opposed to the case of $q\,{=}\,0$ \cite{Sm,IS14a}.

For the excited states with $\ell\,{>}\,0$\,, Casimirs \p{C2} and \p{C3} take the values
\bea
    {\mathbf C}_2\,\Phi^{(2q,\ell > 0)} = \left(\beta^2 - \lambda^2\right)\Phi^{(2q,\ell > 0)},\qquad
{\mathbf C}_3\,\Phi^{(2q,\ell > 0)} = \beta\left(\beta^2 - \lambda^2\right)\Phi^{(2q,\ell > 0)},
\eea
where
\bea
    \beta = 2\ell+q+\frac{1}{2}\,,\qquad \lambda = q+\frac{1}{2}\,.
\eea
Thus these states comprise typical representations, with the total dimension $8\lambda\,{=}\,4\left(2q+1\right)$ and equal numbers of bosonic and fermionic states.
The wave functions $\Phi^{(2q,\ell > 0)}$ shows up the full component expansions \eqref{wf-1} involving equal numbers of the bosonic and fermionic states.

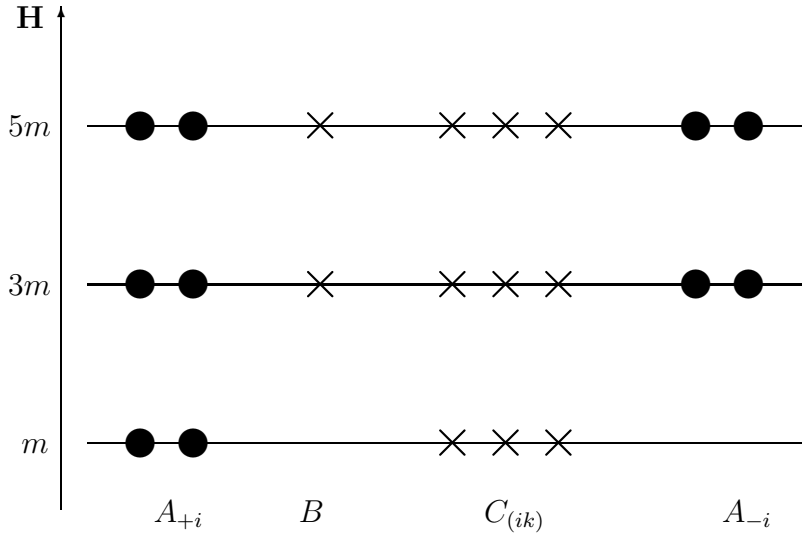
\begin{figure}[ht]
\begin{center}
\begin{picture}(350,200)

\put(20,5){\line(0,1){180}}
\put(20,185){\vector(0,1){10}}

\put(30,30){\line(1,0){270}}
\put(30,90){\line(1,0){270}}
\put(30,150){\line(1,0){270}}

\multiput(50,30)(20,0){2}{\circle*{11}}
\multiput(50,90)(20,0){2}{\circle*{11}}
\multiput(50,150)(20,0){2}{\circle*{11}}

\put(110,85){{\LARGE $\times$}}
\put(110,145){{\LARGE $\times$}}

\multiput(160,25)(20,0){3}{{\LARGE $\times$}}
\multiput(160,85)(20,0){3}{{\LARGE $\times$}}
\multiput(160,145)(20,0){3}{{\LARGE $\times$}}

\multiput(260,90)(20,0){2}{\circle*{11}}
\multiput(260,150)(20,0){2}{\circle*{11}}

\put(5,26){$m$}
\put(0,86){$3m$}
\put(0,146){$5m$}
\put(3,187){${\mathbf H}$}

\put(55,0){$A_{+ i}$}
\put(110,0){$B$}
\put(180,0){$C_{(ik)}$}
\put(270,0){$A_{- i}$}

\end{picture}
\end{center}
\caption{The degeneracy of energy levels for $q\,{=}\,1/2$\,. Circles and crosses indicate bosonic and fermionic states, respectively.}
\label{figure1}
\end{figure}

As an instructive example, let us consider the structure of wave functions for $q\,{=}\,1/2$\,. In this case, the bosonic states $A_{\pm i}$ are doublets,
while the fermionic states $B$ and $C_{(ij)}$ form singlets and triplets. The relevant degeneracy is depicted in Figure \ref{figure1}.

The degeneracy picture for $q\,{>}\,0$ drastically differs from that for $q\,{=}\,0$ found in \cite{Sm,IS14a}.
Here we encounter the vacuum \eqref{Vacuum} comprising $4q\,{+}\,3$ states which belong to an atypical representation of ${\rm SU}(2|1)$,
while for $q\,{=}\,0$ the unique ground state is ${\rm SU}(2|1)$ singlet annihilated by both supercharges.
The states at first excited level for $q\,{=}\,0$ form an atypical fundamental representation of ${\rm SU}(2|1)$ spanned by one bosonic and two fermionic states.
For $q\,{>}\,0$ the states at all excited levels belong to typical ${\rm SU}(2|1)$ representations with equal numbers $[2(2q +1) + 2(2q+1)]$ of bosonic and fermionic components.
Also, the spectrum \eqref{eq-conf-ener-comp} has the level spacing $2m$\,, while the spectrum found in \cite{IS14a} displays the level spacing $m\,$.
This distinction is due to discarding the additional solution \eqref{2sol} for $q\,{>}\,0$\,, while for $q\,{=}\,0$ both the solutions \eqref{2sol} and \eqref{eq-conf-sol}
should be taken into account, which results in doubling of energy levels in the latter case.

The coefficient in the expansion of the full wave function \eqref{wf-1} over spin variables $z^i$ for $\ell\,{>}\,0$ (i.e. for typical representations)
can be interpreted as a chiral superfield ``living'' on ${\rm SU}(2|1)$ quantum superspace, such that it carries $2q$ external symmetrized doublet
indices of ${\rm SU}(2) \subset {\rm SU}(2|1)\,$. Such chiral ${\rm SU}(2|1)$ superfields with $2s$ external symmetrized ${\rm SU}(2)$ indices were considered in \cite{IS-long}.
They describe $d\,{=}\,1$ models obtained by a dimensional reduction from supersymmetric $d\,{=}\,4$ models defined on curved spaces \cite{long}.

In conclusion, let us point out once more that in our model ${\rm SU}(2|1)$ supersymmetry is spontaneously broken
and the vacuum energy \eqref{V-energy} depends on the spin $q$\,. In the ${\rm SU}(2|1)$ mechanics model considered in \cite{long},
the vacuum energy (Casimir energy) depends on some ${\rm U}(1)$ charge, such that the sum of ${\rm U}(1)$ charge and vacuum energy is zero.
This feature is related to the fact that there the vacuum state is ${\rm SU}(2|1)$ singlet and so does not break supersymmetry.
In our case we encounter a spontaneously broken vacuum described by a non-trivial atypical representation of ${\rm SU}(2|1)$, with the spin $q$ as
an intrinsic parameter of this vacuum. For example, in the case $q=1/2$ depicted in Figure \ref{figure1} the vacuum states are represented by the fields $A_{+ i}$ and $C_{(ik)}$ which are
the doublet and the triplet of the internal ${\rm SU}(2)$ group with the generators $\mathbf{I}^i_k$. So, the vacuum is not invariant under the action of the subgroup ${\rm SU}(2)\subset {\rm SU}(2|1)$,
and this ${\rm SU}(2)$ is spontaneously broken too. It is instructive to give explicitly how the supercharges act on the vacuum states in this case. In accord with \p{26u}, the vacuum state $\Phi^{(1,0)}$ is written as
    \begin{eqnarray}
        \Phi^{(1,0)} =  z^k A_{+\,k}\,|0, 3/2\,\rangle + \psi^{(j} z^{k)} C_{(jk)}\,|0, 1/2\,\rangle.
    \end{eqnarray}
It is straightforward to find the action of supercharges on these functions
    \begin{eqnarray}
   && \mathbf{Q}^i z^k A_{+\,k}\,|0, 3/2\,\rangle = -i \psi^{(j} z^{k)}A_{+\,k}|0, 1/2\,\rangle \,,  \quad  \bar{\mathbf{Q}}^i z^k A_{+\,k}\,|0, 3/2\,\rangle  = 0\,,\nn
   && \bar{\mathbf{Q}}_i\psi^{(j} z^{k)} C_{(jk)}\,|0, 1/2\,\rangle = 2im \delta^j_i z^{k}C_{(jk)}|0, 3/2\,\rangle \,, \quad \mathbf{Q}_i\psi^{(j} z^{k)} C_{(jk)}\,|0, 1/2\,\rangle = 0\,.
    \end{eqnarray}
The action of the bosonic generators can also be directly found, using the explicit expressions for them.

\section{Hidden superconformal symmetry}

In this Section we show that the action \eqref{L} is invariant under $\mathcal{N}\,{=}\,4$ superconformal symmetry ${\rm OSp}(4|2)\,$.
Part of it appears as a hidden symmetry of the model. This property is somewhat unexpected, since
the system \eqref{L} includes the mass parameter $m\,$.

This additional symmetry can be easily revealed after passing from the original fermionic fields to the new ones
\bea
    \chi^i := \psi^i e^{-imt},\qquad \bar{\chi}_j := \bar{\psi}_j e^{imt}\,. \label{redef}
\eea
In terms of these new variables the Lagrangian \eqref{L} takes the form
\bea
    {\cal L} &=& \frac{\dot{x}^2}{2}+ \frac{i}{2}\left(\dot{\bar z}_k z^k -
{\bar z}_k \dot z^k\right) + \frac{i}{2}\left(\bar{\chi}_k \dot\chi^k -\dot{\bar\chi}_k \chi^k \right)-\frac{\left(z^{k}\bar{z}_k\right)^2}{8x^2}
-\frac{\chi^{(i}\bar{\chi}^{j)} z_{(i} \bar z_{j)}}{x^2}\nn
    &&-\,\frac{m^2x^2}{2}+A\left(z^k\bar{z}_k-c\right). \label{L-red}
\eea
In contrast to the Lagrangian \eqref{L}, the Lagrangian \eqref{L-red} does not contain mass terms for the fermionic fields.

Obviously, the Lagrangian \eqref{L-red} remains invariant under the ${\rm SU}(2|1)$ transformations \eqref{tr},
which, after the redefinition \p{redef}, acquire the following $t$-dependent form
\bea
    &&\delta x = -\,\epsilon_{+\,k}\chi^k e^{imt} +\bar{\epsilon}^k_{+} \bar{\chi}_k e^{-imt}, \nn
    &&\delta \chi^i = \left[\bar{\epsilon}^i_{+} \left(i\dot{x}-m\,x\right)+
\frac{\bar{\epsilon}_{+\,k} z^{(i}z^{k)}}{x}\right]e^{-imt},\nn
    &&\delta \bar{\chi}_j =\left[-\,\epsilon_{+\,j}\left(i\dot{x}+m\,x\right) + \frac{\epsilon^k_{+} z_{(j}z_{k)}}{x}
\right]e^{imt},\nn
    &&\delta z^{i} = \frac{z_{k}}{x}\left[\epsilon^{(i}_{+}\chi^{k)}e^{imt}+\bar{\epsilon}^{(i}_{+\,k}\bar{\chi}^{k)}e^{-imt}\right],\nn
    &&\delta \bar{z}_{j} =-\,\frac{\bar{z}^{k}}{x}\left[\epsilon_{+\,(j}\chi_{k)}e^{imt}+\bar{\epsilon}_{+\,(j}\bar{\chi}_{k)}e^{-imt}\right],\label{tr+}
\eea
where
\bea
    \epsilon_{+\,k} := \epsilon_k\,,\qquad \bar{\epsilon}^k_{+} := \bar{\epsilon}^k.
\eea
We see that the Lagrangian is an even function of $m$, it depends only on $m^2$. Hence, it is also invariant
under the new type of $\epsilon_{-\,k}$ and $\bar{\epsilon}^k_{-}$ transformations
\bea
    &&\delta x = -\,\epsilon_{-\,k}\chi^k e^{-imt} +\bar{\epsilon}^k_{-} \bar{\chi}_k e^{imt}, \nn
    &&\delta \chi^i = \left[\bar{\epsilon}^i_{-} \left(i\dot{x}+m\,x\right)+
\frac{\bar{\epsilon}_{-\,k} z^{(i}z^{k)}}{x}\right]e^{imt},\nn
    &&\delta \bar{\chi}_j =\left[-\,\epsilon_{+\,j}\left(i\dot{x}-m\,x\right) + \frac{\epsilon^k_{+} z_{(j}z_{k)}}{x}
\right]e^{-imt},\nn
    &&\delta z^{i} = \frac{z_{k}}{x}\left[\epsilon^{(i}_{-}\chi^{k)}e^{-imt}+\bar{\epsilon}^{(i}_{-\,k}\bar{\chi}^{k)}e^{imt}\right],\nn
    &&\delta \bar{z}_{j} =-\,\frac{\bar{z}^{k}}{x}\left[\epsilon_{-\,(j}\chi_{k)}e^{-imt}+\bar{\epsilon}_{-\,(j}\bar{\chi}_{k)}e^{imt}\right],\label{tr-}
\eea
which are obtained from \eqref{tr+} just through the substitution $m\rightarrow -m$.
This feature is typical for the trigonometric type of superconformal symmetry \cite{ISTconf}.
As we will see below, the closure of these two types of transformations gives the superconformal group $D(2,1;-1/2)\cong {\rm OSp}(4|2)$, where the superconformal Hamiltonian is defined as
\be\label{ca-H}
    {\cal H} := H-2mF.
\ee
Such a redefinition of the Hamiltonian, together with passing to the fermionic fields \eqref{redef},
ensure a proper embedding of the extended superalgebra $\hat{su}(2|1) = su(2|1) +\!\!\!\!\!\!\!\supset u(1)$ into the superconformal algebra $osp(4|2)$.

\subsection{Hamiltonian analysis and extended superalgebra}

The Lagrangian \eqref{L-red} yields, modulo the constraint-generating term $A\left(z^k\bar{z}_k-c\right)$, just the conformal Hamiltonian \p{ca-H} as the canonical one:
\be
    {\cal H} = \frac12\left(p^2 + m^2 x^2\right) - \frac{S^{(ij)}S_{(ij)}}{4x^2}
+ \frac{\chi^{(i}\bar{\chi}^{j)} S_{(ij)}}{x^2}\,.\label{conf-H}
\ee
Like the Lagrangian \eqref{L-red}, the Hamiltonian \eqref{conf-H}, as opposed to \eqref{H}, does not contain mass terms for the fermionic fields.

The canonical Dirac brackets of the variables entering the system \eqref{L-red} are given by
\be
    \left\{x, p \right\}^*=1\,,\qquad \left\{z^i, \bar z_j \right\}^* = i\delta_{j}^i\,,\qquad
\left\{\chi^i, \bar{\chi}_{j}\right\}^* = -\,i\delta_{j}^i\,.\label{brackets-new}
\ee
The transformations \eqref{tr+} and \eqref{tr-} are generated by the Noether supercharges
\bea
    Q^i_+ = \left[\left(p-imx\right)\chi^i + \frac{i}{x}\,S^{(ik)}\chi_{k}\right]e^{imt}\,,\qquad
    \bar{Q}_{+}{}_{j} = \left[\left(p+imx\right)\bar{\chi}_j-\frac{i}{x}\,S_{(jk)}\bar{\chi}^{k}\right]e^{-imt}\,,\label{Q+}\\
    Q^i_- = \left[\left(p+imx\right)\chi^i + \frac{i}{x}\,S^{(ik)}\chi_{k}\right]e^{-imt}\,,\qquad
    \bar{Q}_{-}{}_{j} = \left[\left(p-imx\right)\bar{\chi}_j-\frac{i}{x}\,S_{(jk)}\bar{\chi}^{k}\right]e^{imt}\,.\label{Q-}
\eea
Computing the closure of these supercharges
\bea\label{Q-Q}
&&\left\{Q^i_{\pm}, \bar{Q}_{\pm}{}_j \right\}^*=-\,2i\delta^i_j\,{\cal H}\mp 2im\left(I^i_j + \delta^i_j F \right)\,,\nn
&&\left\{Q^i_{-}, \bar{Q}_{+}{}_j \right\}^*=-\,2i\delta^i_j \,{\cal N}\,,\qquad
\left\{Q^i_{+}, \bar{Q}_{-}{}_j \right\}^*=-\,2i\delta^i_j \,\bar{\cal N}\,,\nn
&&\left\{Q^i_{+}, Q^k_{-}\right\}^*=2im\epsilon^{ik}\,G\,,\qquad
\left\{\bar{Q}_{+}{}_j, \bar{Q}_{-}{}_k \right\}^*=-\,2im\epsilon_{jk}\,\bar{G}\,,
\eea
we recover the bosonic generators ${\cal H}$, $I^i_j$, $F$ defined in \eqref{conf-H}, \eqref{b-gen}, along with the new bosonic
generators
\bea
    &&G = \frac{1}{2}\,\chi^k\chi_k\,, \qquad \bar{G} = \frac{1}{2}\,\bar{\chi}_k\bar{\chi}^k,\nn
    &&{\cal N} = e^{-2imt}\left[\frac{1}{2}\,\left(p+imx\right)^2-\frac{S^{(ij)}S_{(ij)}}{4x^2}+ \frac{\chi^{(i}\bar{\chi}^{j)} S_{(ij)}}{x^2}\right],\nn
    &&\bar{\cal N} = e^{2imt}\left[\frac{1}{2}\,\left(p-imx\right)^2-\frac{S^{(ij)}S_{(ij)}}{4x^2}+ \frac{\chi^{(i}\bar{\chi}^{j)} S_{(ij)}}{x^2}\right].\label{new-b-gen}
\eea
The bosonic generators \eqref{b-gen}, \eqref{conf-H} and \eqref{new-b-gen} satisfy the following Dirac bracket algebra
\bea\label{B-B}
    &&\left\{I^i_k, I^j_l \right\}^*=i\left(\delta^i_l I^j_k - \delta^j_k I^i_l\right),\nn
    &&\left\{G, \bar{G}\right\}^*=-\,2iF,\qquad \left\{F, G\right\}^*=-\,iG,\qquad\left\{F, \bar{G}\right\}^*=i\bar{G},\nn
    &&\left\{{\cal N}, \bar{\cal N}\right\}^*=4im{\cal H},\qquad
\left\{{\cal H}, {\cal N}\right\}^*=-\,2im{\cal N},\qquad\left\{{\cal H}, \bar{\cal N}\right\}^*=2im\bar{\cal N}.
\eea
It is a direct sum of three mutually commuting algebras
\be\label{3-b-su}
I^i_j\oplus \left(G,\bar{G},F\right)\oplus \left({\cal H},{\cal N},\bar{\cal N}\right)
\quad = \quad
su_R(2)\oplus su_L(2) \oplus so(2,1).
\ee
The algebra $so(2,1)$ with the generators ${\cal H}$, ${\cal N}$, $\bar{\cal N}$ is just $d\,{=}\,1$ conformal algebra.
The supercharges \eqref{Q+} and \eqref{Q-} transform under the generators \eqref{3-b-su} as
\bea
&&\left\{I^i_j, Q^k_{\pm} \right\}^*=-\,i\left(\delta_j^k Q^i_{\pm} - \frac{\delta_j^i}{2}\,Q^k_{\pm}\right),
\qquad \left\{I^i_j, \bar{Q}_{\pm}{}_k \right\}^*=
i\left(\delta_k^i\bar{Q}_{\pm}{}_j-\frac{\delta_j^i}{2}\,\bar{Q}_{\pm}{}_{k}\right),\nn
&&\left\{F, Q^i_{\pm} \right\}^*=-\frac{i}{2}\,Q^i_{\pm}\,,\qquad
    \left\{F, \bar{Q}_{\pm}{}_j \right\}^*=\frac{i}{2}\,\bar{Q}_{\pm}{}_j\,,\nn
&&\left\{\bar{G}, Q^i_{\pm} \right\}^*=i\epsilon^{ik}\bar{Q}_{\pm}{}_k\,,\qquad
    \left\{G, \bar{Q}_{\pm}{}_j \right\}^*=i\epsilon_{jk}Q^k_{\mp}\,,\nn
&&\left\{{\cal H}, Q^i_{\pm} \right\}^*=\pm\,im Q^i_{\pm}\,,\qquad
    \left\{{\cal H}, \bar{Q}_{\pm}{}_j \right\}^*=\mp\,im\bar{Q}_{\pm}{}_j\,,\nn
&&\left\{{\cal N}, Q^i_{+}\right\}^*=2imQ^i_{-}\,,\qquad
    \left\{\bar{\cal N}, \bar{Q}_+{}_j \right\}^*=-\,2im\bar{Q}_-{}_j\,,\nn
&&\left\{\bar{\cal N}, Q^i_{-}\right\}^*=-\,2im Q^i_{+}\,,\qquad
    \left\{{\cal N}, \bar{Q}_-{}_j \right\}^*=2im\bar{Q}_+{}_j\,.\label{conf-algebra}
\eea
All the remaining commutators between the supercharges \eqref{Q+}, \eqref{Q-} and the generators \eqref{3-b-su} are vanishing.

The superalgebra defined by (anti)commutators \eqref{Q-Q}, \eqref{B-B}, and \eqref{conf-algebra}
is the superconformal algebra $osp(4|2)$ in ``AdS basis''. The parameter $m$ appearing in {\cal r.h.s.} of  \eqref{Q-Q}, \eqref{B-B}, \eqref{conf-algebra}
plays the role of the inverse radius of AdS$_2 \sim {\rm O}(2,1)/{\rm O}(1,1)\,$.

The standard form of the $osp(4|2)$ superalgebra \cite{FIL09,FIL10,FIL12} is recovered after passing to the new basis
\bea
    && {\cal Q}^i:=-\frac{1}{2}\left(Q^i_{+} + Q^i_{-}\right),\qquad
    \bar{\cal Q}_j:=-\frac{1}{2}\left(\bar{Q}_+{}_j +  \bar{Q}_-{}_j\right),\nn
    && {\cal S}^{i}:=\frac{i}{2m}\left(Q^i_{+} - Q^i_{-}\right),\qquad
    \bar{\cal S}_j:=-\frac{i}{2m}\left(\bar{Q}_+{}_j - \bar{Q}_-{}_j \right),\nn
    && \tilde{K}:=\frac{1}{2m^2}\left[{\cal H} - \frac{1}{2}\left({\cal N}+\bar{\cal N}\right)\right],\qquad
    \tilde{H} :=
    \frac{1}{2}\left[{\cal H} + \frac{1}{2}\left({\cal N}+\bar{\cal N}\right)\right],\nn
    && \tilde{D}:=\frac{i}{4m}\left({\cal N}-\bar{\cal N}\right),\qquad m \neq 0\,.\label{gener-new}
\eea
In terms of the redefined generators, the Dirac bracket algebra \eqref{Q-Q}, \eqref{B-B} and \eqref{conf-algebra} is rewritten as
\bea
    &&\left\{{\cal Q}^i, \bar{\cal Q}_j\right\}^*=-\,2i\delta^i_j\tilde{H}\,,\qquad \left\{{\cal S}^i, \bar{\cal S}_j\right\}^*=-\,2i\delta^i_j\tilde{K}\,,\nn
    &&\left\{{\cal Q}^i, \bar{\cal S}_j\right\}^*=-\,2i\delta^i_j\tilde{D}+I^i_j + \delta^i_j F\,,\qquad
    \left\{{\cal S}^i, \bar{\cal Q}_j\right\}^*=-\,2i\delta^i_j\tilde{D}-I^i_j - \delta^i_j F\,,\nn
    &&\left\{{\cal Q}^i, {\cal S}^k\right\}^*=-\,\epsilon^{ik}G\,,\qquad
    \left\{\bar{\cal Q}_j,\bar{\cal S}_k\right\}^*=-\,\epsilon_{jk}\bar{G}\,,\label{osp42-qq}
\eea
\bea
    &&\left\{I^i_k, I^j_l \right\}^*=i\left(\delta^i_l I^j_k - \delta^j_k I^i_l\right),\nn
    &&\left\{G, \bar{G}\right\}^*=-\,2iF,\quad \left\{F, G\right\}^*=-\,iG,\quad\left\{F, \bar{G}\right\}^*=i\bar{G}\,,\nn
    &&\left\{\tilde{H}, \tilde{K}\right\}^*=2\tilde{D}\,,\quad \left\{\tilde{D}, \tilde{H}\right\}^*=-\,\tilde{H}\,,\quad\left\{\tilde{D}, \tilde{K}\right\}^*=\tilde{K}\,,\label{osp42-bb}
\eea
\bea
    &&\left\{I^i_j, {\cal Q}^k \right\}^*=-\,i\left(\delta_j^k {\cal Q}^i - \frac{1}{2}\,\delta_j^i{\cal Q}^k\right),
\quad \left\{I^i_j, \bar{\cal Q}_k \right\}^*=
i\left(\delta_k^i\bar{\cal Q}_j-\frac{1}{2}\,\delta_j^i\bar{\cal Q}_{k}\right),\nn
    &&\left\{I^i_j, {\cal S}^k \right\}^*=-\,i\left(\delta_j^k {\cal S}^i - \frac{1}{2}\,\delta_j^i{\cal S}^k\right),
\quad \left\{I^i_j, \bar{\cal S}_k \right\}^*=
i\left(\delta_k^i\bar{\cal S}_j-\frac{1}{2}\,\delta_j^i\bar{\cal S}_{k}\right),\nn
    &&\left\{F, {\cal Q}^i\right\}^*=-\frac{i}{2}\,{\cal Q}^i,\quad
    \left\{F, \bar{\cal Q}_j \right\}^*=\frac{i}{2}\,\bar{\cal Q}_j\,,\qquad
    \left\{F, {\cal S}^i\right\}^*=-\frac{i}{2}\,{\cal S}^i,\quad
    \left\{F, \bar{\cal S}_j \right\}^*=\frac{i}{2}\,\bar{\cal S}_j\,,\nn
    &&\left\{\bar{G},{\cal Q}^i\right\}^*=i\bar{\cal Q}^i\,,\quad
    \left\{G, \bar{\cal Q}_j\right\}^*=i{\cal Q}_j\,,\qquad
    \left\{\bar{G},{\cal S}^i\right\}^*=i\bar{\cal S}^i\,,\qquad
    \left\{G, \bar{\cal S}_j\right\}^*=i{\cal S}_j\,,\nn
    &&\left\{\tilde{D}, {\cal Q}^i\right\}^*= -\frac12\,{\cal Q}^i\,,\quad
    \left\{\tilde{D}, \bar{\cal Q}_j\right\}^*=-\frac12\,\bar{\cal Q}_j\,,\quad
    \left\{\tilde{D}, {\cal S}^i\right\}^*= \frac12\,{\cal S}^i\,,\quad
    \left\{\tilde{D}, \bar{\cal S}_j\right\}^*=\frac12\,\bar{\cal S}_j
\,,\nn
    &&\left\{\tilde{H}, {\cal S}^i\right\}^*= {\cal Q}^i\,,\quad
    \left\{\tilde{H}, \bar{\cal S}_j\right\}^*=\bar{\cal Q}_j\,,\qquad
    \left\{\tilde{K}, {\cal Q}^i\right\}^*= -\,{\cal S}^i\,,\quad
    \left\{\tilde{K}, \bar{\cal Q}_j\right\}^*=-\,\bar{\cal S}_j\,.\label{osp42-bq}
\eea
This is  the standard form of the superalgebra $osp(4|2)\,$.

In contradistinction to the relations in the basis  \eqref{Q-Q}, \eqref{B-B} and \eqref{conf-algebra}, the relations \eqref{osp42-qq}, \eqref{osp42-bb} and \eqref{osp42-bq}
do not involve the parameter $m\,$, though it
is still present in the expressions for almost all generators [see \eqref{conf-H}, \eqref{Q+}, \eqref{Q-} and \eqref{new-b-gen}]. Now it is easy to check that
the new generators defined in \p{gener-new} are non-singular at $m=0$ and so one can consider the ``pure conformal'' limit $m\,{\to}\,0$
in these expressions without affecting the (anti)commutation relations \eqref{osp42-qq} - \eqref{osp42-bq}.
In this limit, the generators \eqref{b-gen}, \eqref{gener-new}, \eqref{new-b-gen} coincide with those given in \cite{FIL09}.
In particular, the conformal generators acquire the ``parabolic'' form
\bea
    \tilde{H} = \frac{p^2}{2} - \frac{S^{(ij)}S_{(ij)}}{4x^2}
+ \frac{\chi^{(i}\bar{\chi}^{j)} S_{(ij)}}{x^2},\qquad
    \tilde{D} = -\frac{1}{2}\,xp + t \tilde{H},\qquad
    \tilde{K} = \frac{x^2}{2}-txp + t^2\tilde{H}.
\eea
The parabolic realization of ${\rm OSp}(4|2)$ leaves invariant the $m=0$ limit of the action \p{L}. It is just
the action considered in \cite{FIL09}.

\subsection{Superconformal symmetry of the spectrum}

Here we consider the implications of the hidden superconformal symmetry for the spectrum
derived in Sect.\,3.2.

Quantum counterparts of the superconformal odd generators
\eqref{Q+}, \eqref{Q-} and \eqref{gener-new} are uniquely determined:
\bea
&& \mbox{\textsf{Q}}^i=-\frac{1}{2}\left(\mathbf{Q}^i_{+} + \mathbf{Q}^i_{-}\right),\qquad
\bar{\mbox{\textsf{Q}}}_j=-\frac{1}{2}\left(\bar{\mathbf{Q}}_+{}_j +  \bar{\mathbf{Q}}_-{}_j\right),\nn
&& \mbox{\textsf{S}}^{i}=\frac{i}{2m}\left(\mathbf{Q}^i_{+} - \mathbf{Q}^i_{-}\right),\qquad
\bar{\mbox{\textsf{S}}}_j=-\frac{i}{2m}\left(\bar{\mathbf{Q}}_+{}_j - \bar{\mathbf{Q}}_-{}_j \right),\label{Q-new-q}
\eea
where
\bea
\mathbf{Q}^i_+ = e^{imt}\left[\left(\mathbf{p}-im\mathbf{x}\right){\bm\chi}^i + \frac{i}{\mathbf{x}}\,\mathbf{S}^{(ik)}{\bm\chi}_{k}\right],\quad
\bar{\mathbf{Q}}_{+}{}_{j} = e^{-imt}\left[\left(\mathbf{p}+im\mathbf{x}\right)\bar{\bm\chi}_j-
\frac{i}{\mathbf{x}}\,\mathbf{S}_{(jk)}\bar{\bm\chi}^{k}\right],\label{Q+-q}\\
\mathbf{Q}^i_- = e^{-imt}\left[\left(\mathbf{p}+im\mathbf{x}\right){\bm\chi}^i + \frac{i}{\mathbf{x}}\,\mathbf{S}^{(ik)}{\bm\chi}_{k}\right],\quad
\bar{\mathbf{Q}}_{-}{}_{j} = e^{imt}\left[\left(\mathbf{p}-im\mathbf{x}\right)\bar{\bm\chi}_j-
\frac{i}{\mathbf{x}}\,\mathbf{S}_{(jk)}\bar{\bm\chi}^{k}\right].\label{Q--q}
\eea
Their anticommutators,
\bea
&&\left\{\mbox{\textsf{Q}}^i, \bar{\mbox{\textsf{Q}}}_j\right\} =2\delta^i_j\,\tilde{\mathbf{H}}\,,\qquad
\left\{\mbox{\textsf{S}}^i, \bar{\mbox{\textsf{S}}}_j\right\} =2\delta^i_j\,\tilde{\mathbf{K}}\,,\nn
&&\left\{\mbox{\textsf{Q}}^i, \bar{\mbox{\textsf{S}}}_j\right\} =2\delta^i_j\,\tilde{\mathbf{D}}+
i\,\mathbf{I}^i_j + i\delta^i_j\, \mathbf{F}\,,\qquad
\left\{\mbox{\textsf{S}}^i, \bar{\mbox{\textsf{Q}}}_j\right\} =2\delta^i_j\,\tilde{\mathbf{D}}-
i\,\mathbf{I}^i_j - i\delta^i_j \,\mathbf{F}\,,\nn
&&\left\{\mbox{\textsf{Q}}^i, \mbox{\textsf{S}}^k\right\} =-\,i\epsilon^{ik}\,\mathbf{G}\,,\qquad
\left\{\bar{\mbox{\textsf{Q}}}_j,\bar{\mbox{\textsf{S}}}_k\right\} =-\,i\epsilon_{jk}\,\bar{\mathbf{G}}\,, \label{osp42-qq-q}
\eea
give the expressions for the quantum even generators
\be
{\mathbf{F}}=\frac{1}{4}\,\left[{\bm{\chi}}^k,\bar{\bm{\chi}}_k\right]\,,\qquad
{\mathbf{G}}=\frac{1}{2}\,{\bm{\chi}}^k{\bm{\chi}}_k \,,\qquad
\bar{\mathbf{G}}=\frac{1}{2}\,\bar{\bm{\chi}}_k\bar{\bm{\chi}}^k\,,\label{su2r-q}
\ee
\be
{\mathbf I}^i_k=\epsilon_{kj}\left({\mathbf S}^{(ij)}+ {\bm{\chi}}^{(i}\bar{\bm{\chi}}^{j)}\right)\,,\label{su2l-q}
\ee
\bea
\tilde{\mathbf{H}}&=&
\frac{1}{2}\left[\mbox{\textbf{\textit{H}}} + \frac{1}{2}\left(\mbox{\textbf{\textit{N}}}+\bar{\mbox{\textbf{\textit{N}}}}\right)\right],\nn
\tilde{\mathbf{K}}&=&
\frac{1}{2m^2}\left[\mbox{\textbf{\textit{H}}} - \frac{1}{2}\left(\mbox{\textbf{\textit{N}}}+\bar{\mbox{\textbf{\textit{N}}}}\right)\right],\nn
\tilde{\mathbf{D}}&=&\frac{i}{4m}\left(\mbox{\textbf{\textit{N}}}-\bar{\mbox{\textbf{\textit{N}}}}\right). \label{sl2-new-q}
\eea
Here
\bea
\mbox{\textbf{\textit{H}}}&=&\frac12\left(\mathbf{p}^2 + m^2 \mathbf{x}^2\right) - \frac{\mathbf{S}^{(ij)}\mathbf{S}_{ij}}{4\mathbf{x}^2}
+ \frac{{\bm\chi}^{(i}\bar{{\bm\chi}}^{j)} \mathbf{S}_{ij}}{\mathbf{x}^2}\,,\label{Hconf5} \\
\mbox{\textbf{\textit{N}}}&=&e^{-2imt}\left[\frac{1}{2}\,\left(\mathbf{p}+im\mathbf{x}\right)^2-\frac{\mathbf{S}^{(ij)}\mathbf{S}_{ij}}{4\mathbf{x}^2}
+ \frac{{\bm{\chi}}^{(i}\bar{\bm{\chi}}^{j)} \mathbf{S}_{ij}}{\mathbf{x}^2}\right],\nn
\bar{\mbox{\textbf{\textit{N}}}}&=&e^{2imt}\left[\frac{1}{2}\,\left(\mathbf{p}-im\mathbf{x}\right)^2
-\frac{\mathbf{S}^{(ij)}\mathbf{S}_{ij}}{4\mathbf{x}^2}+ \frac{{\bm{\chi}}^{(i}\bar{\bm{\chi}}^{j)} \mathbf{S}_{ij}}{\mathbf{x}^2}\right]\label{sl2r-new-q}
\eea
are the quantum counterparts of the classical quantities \eqref{conf-H} and \eqref{new-b-gen}.
The generators \eqref{Q+-q} and \eqref{Q--q} will go over to the trigonometric realization of the ${\rm OSp}(4|2)$
generators given in \cite{CHolTop}, if we pass to the case without spin degrees of freedom.
Note that the spectrum and the energy of the vacuum states in the models of the $D(2,1;\alpha)$ superconformal
mechanics is known to exhibit an explicit  dependence on $\alpha$ \cite{FIL10, CHolTop}. The vacuum energy of the ``conformal'' Hamiltonian \p{Hconf5} including
the case of $q=0$ is also defined by the expression \eqref{V-energy}.
This vacuum energy for the spinless system ({\it i.e.} for $q=0$) coincides with the one given in \cite{CHolTop} (for $\alpha = -1/2$), if we choose
$m=1/2\,$.

The second-order Casimir operator of ${\rm OSp}(4|2)=D(2,1;-\frac12)$ is defined by the
following expression \cite{Je,FIL09,FIL12}\footnote{The $\mathrm{OSp}(4|2)$ generators in our notation are related to those
quoted in the papers \cite{FIL09,FIL12}, through the substitutions
$\mathbf{I} \to i\mathbf{I}$, ${\mbox{\textsf{Q}}} \to \bar{\mbox{\textsf{Q}}}$, ${\mbox{\textsf{S}}} \to \bar{\mbox{\textsf{S}}}$.}
\begin{equation}\label{qu-Cas}
\mathbf{C}^{\,\prime}_2=\frac{1}{2}\,\left\{\tilde{\mathbf{H}},\tilde{\mathbf{K}} \right\}-\tilde{\mathbf{D}}^2 \,-\,
\frac{1}{4}\,\left\{{\mathbf{G}},\bar{\mathbf{G}} \right\}-\frac{1}{2}\,{\mathbf{F}}^2
\,+\,\frac{1}{4}\, \mathbf{I}^{ik}\mathbf{I}_{ik} \,-\,
\frac{i}{4}\, [{\mbox{\textsf{Q}}}^{i},\bar {\mbox{\textsf{S}}}_{i}] -
\frac{i}{4}\, [\bar {\mbox{\textsf{Q}}}_{i}, {\mbox{\textsf{S}}}^{i}]\,.
\end{equation}
Here $\frac{1}{2}\,\left\{\tilde{\mathbf{H}},\tilde{\mathbf{K}} \right\}{-}\,\tilde{\mathbf{D}}^2$ is the Casimir operator
of $o(2,1)$, whereas
$\frac{1}{2}\,\left\{{\mathbf{G}},\bar{\mathbf{G}} \right\}{+}\, {\mathbf{F}}^2$ and
$\frac{1}{2}\, \mathbf{I}^{ik}\mathbf{I}_{ik}$ are the Casimirs of $su_R(2)$ and $su_L(2)$.
Substituting the explicit expressions  \eqref{Q+-q}, \eqref{Q--q}, \eqref{su2r-q} -  \eqref{sl2r-new-q} for the generators,
we obtain
\begin{eqnarray}
&&\frac{1}{2}\,\left\{\tilde{\mathbf{H}},\tilde{\mathbf{K}} \right\}{-}\,\tilde{\mathbf{D}}^2=
-\frac{1}{8}\, \mathbf{S}^{(ij)}\mathbf{S}_{ij}  +
\frac{1}{2}\,{\bm{\chi}}^{(i}\bar{\bm{\chi}}^{j)} \mathbf{S}_{ij} -
\frac{3}{16}\,, \nonumber
\\
&&\frac{1}{2}\,\left\{{\mathbf{G}},\bar{\mathbf{G}} \right\}{+}\, {\mathbf{F}}^2=
\frac{3}{8}\,\left({\bm{\chi}}^i{\bm{\chi}}_i\, \bar{\bm{\chi}}_k\bar{\bm{\chi}}^k -2
{\bm{\chi}}^i \bar{\bm{\chi}}_i \right)+ \frac{3}{4}
, \nonumber
\\
&&\frac{1}{2}\, \mathbf{I}^{ik}\mathbf{I}_{ik}=
\frac12\,\mathbf{S}^{(ij)}\mathbf{S}_{ij} +
{\bm{\chi}}^{(i}\bar{\bm{\chi}}^{j)} \mathbf{S}_{ij} +
\frac{3}{8}\,\left({\bm{\chi}}^i{\bm{\chi}}_i\, \bar{\bm{\chi}}_k\bar{\bm{\chi}}^k -2
{\bm{\chi}}^i \bar{\bm{\chi}}_i \right) \nonumber
\end{eqnarray}
and
\begin{equation}\label{QQ-quCas}
\frac{i}{4}\, [{\mbox{\textsf{Q}}}^{i},\bar {\mbox{\textsf{S}}}_{i}] +
\frac{i}{4}\, [\bar {\mbox{\textsf{Q}}}_{i}, {\mbox{\textsf{S}}}^{i}]
= -\frac{1}{8m}\, [\mathbf{Q}^i_+ , \bar{\mathbf{Q}}_{+}{}_{i}] +
\frac{1}{8m}\, [\mathbf{Q}^i_- , \bar{\mathbf{Q}}_{-}{}_{i}]
= {\bm{\chi}}^{(i}\bar{\bm{\chi}}^{j)} \mathbf{S}_{ij}
- \frac{1}{2}\,.
\end{equation}
Then the ${\rm OSp}(4|2)$ Casimir operator \eqref{qu-Cas} for the specific system we are considering is equal to
\begin{equation}\label{qu-Cas-12}
\mathbf{C}^{\,\prime}_2= \frac{1}{8}\, \left[\mathbf{S}^{(ij)}\mathbf{S}_{ij}-\frac12\right]
\,.
\end{equation}
Taking into account eq. (\ref{iq-ST}) and the constraint (\ref{T-constr-wf}) for the physical states,
we find that the Casimir operator of the superconformal group ${\rm OSp}(4|2)$ on these states
takes the fixed value
\begin{equation}\label{qu-Cas-12-f}
\mathbf{C}^{\,\prime}_2= -\frac{(2q+1)^2}{16}
\,.
\end{equation}

Thus we come to the following important result: all physical states of the model with ${\rm SU}(2|1)$ supersymmetry,
obtained in the Sect.\,3.2, belong to an irreducible representation of the ${\cal N}{=}\,4$ superconformal group ${\rm OSp}(4|2)\,$.
The states with a fixed eigenvalue of the Hamiltonian of the deformed mechanics $\mathbf{H}$ presented in Sect.\,3.2,
form an irreducible multiplet of  the supergroup $\widehat{{\rm SU}}(2|1) \subset {\rm OSp}(4|2)\,$.
The operators $\mathbf{Q}^i_-$ and $\bar{\mathbf{Q}}_-{}_i$, defined in (\ref{Q--q}),
mix physical states associated with different energy levels, i.e. mix different ${\rm SU}(2|1)$ (or $\widehat{{\rm SU}}(2|1)$)  multiplets of the wave functions. It is straightforward
to check that no any smaller subspace invariant under this action can be singled out in the Hilbert space spanned by all physical states.

\begin{figure}[ht]
\begin{center}
\begin{picture}(350,220)

\put(20,5){\line(0,1){200}}
\put(20,205){\vector(0,1){10}}

\put(30,30){\line(1,0){270}}
\put(30,60){\line(1,0){270}}
\put(30,90){\line(1,0){270}}
\put(30,120){\line(1,0){270}}
\put(30,150){\line(1,0){270}}
\put(30,180){\line(1,0){270}}

\multiput(50,60)(20,0){2}{\circle*{11}}
\multiput(50,120)(20,0){2}{\circle*{11}}
\multiput(50,180)(20,0){2}{\circle*{11}}

\put(110,85){{\LARGE $\times$}}
\put(110,145){{\LARGE $\times$}}

\multiput(160,25)(20,0){3}{{\LARGE $\times$}}
\multiput(160,85)(20,0){3}{{\LARGE $\times$}}
\multiput(160,145)(20,0){3}{{\LARGE $\times$}}

\multiput(260,60)(20,0){2}{\circle*{11}}
\multiput(260,120)(20,0){2}{\circle*{11}}
\multiput(260,180)(20,0){2}{\circle*{11}}

\put(80,55){\line(3,-1){45}}
\put(125,40){\vector(3,-1){10}}
\put(130,41){\tiny ${\bf Q}^i_{+}$}
\put(135,34){\line(-3,1){45}}
\put(90,49){\vector(-3,1){10}}
\put(75,45){\tiny $\bar{\bf Q}_{+j}$}

\put(80,115){\line(3,-1){45}}
\put(125,100){\vector(3,-1){10}}
\put(130,101){\tiny ${\bf Q}^i_{+}$}
\put(135,94){\line(-3,1){45}}
\put(90,109){\vector(-3,1){10}}
\put(75,105){\tiny $\bar{\bf Q}_{+j}$}

\put(80,175){\line(3,-1){45}}
\put(125,160){\vector(3,-1){10}}
\put(130,161){\tiny ${\bf Q}^i_{+}$}
\put(135,154){\line(-3,1){45}}
\put(90,169){\vector(-3,1){10}}
\put(75,165){\tiny $\bar{\bf Q}_{+j}$}

\put(80,62){\line(3,1){10}}
\put(95,67){\line(3,1){10}}
\put(110,72){\line(3,1){10}}
\put(125,77){\vector(3,1){10}}
\put(77,74){\tiny $\bar{\bf Q}_{-j}$}
\put(105,74){\line(-3,-1){10}}
\put(120,79){\line(-3,-1){10}}
\put(135,84){\line(-3,-1){10}}
\put(90,69){\vector(-3,-1){10}}
\put(130,72){\tiny ${\bf Q}^i_{-}$}

\put(80,122){\line(3,1){10}}
\put(95,127){\line(3,1){10}}
\put(110,132){\line(3,1){10}}
\put(125,137){\vector(3,1){10}}
\put(77,134){\tiny $\bar{\bf Q}_{-j}$}
\put(105,134){\line(-3,-1){10}}
\put(120,139){\line(-3,-1){10}}
\put(135,144){\line(-3,-1){10}}
\put(90,129){\vector(-3,-1){10}}
\put(130,132){\tiny ${\bf Q}^i_{-}$}

\put(80,182){\line(3,1){10}}
\put(95,187){\line(3,1){10}}
\put(110,192){\line(3,1){10}}
\put(125,197){\vector(3,1){10}}
\put(77,194){\tiny $\bar{\bf Q}_{-j}$}
\put(105,194){\line(-3,-1){10}}
\put(120,199){\line(-3,-1){10}}
\put(135,204){\line(-3,-1){10}}
\put(90,189){\vector(-3,-1){10}}
\put(130,192){\tiny ${\bf Q}^i_{-}$}

\put(197,85){\line(3,-1){45}}
\put(242,70){\vector(3,-1){10}}
\put(243,73){\tiny ${\bf Q}^i_{+}$}
\put(251,64){\line(-3,1){45}}
\put(206,79){\vector(-3,1){10}}
\put(190,73){\tiny $\bar{\bf Q}_{+j}$}

\put(197,145){\line(3,-1){45}}
\put(242,130){\vector(3,-1){10}}
\put(243,133){\tiny ${\bf Q}^i_{+}$}
\put(251,124){\line(-3,1){45}}
\put(206,139){\vector(-3,1){10}}
\put(190,133){\tiny $\bar{\bf Q}_{+j}$}

\put(197,205){\line(3,-1){45}}
\put(242,190){\vector(3,-1){10}}
\put(243,193){\tiny ${\bf Q}^i_{+}$}
\put(251,184){\line(-3,1){45}}
\put(206,199){\vector(-3,1){10}}
\put(190,193){\tiny $\bar{\bf Q}_{+j}$}

\put(272,53){\line(-3,-1){10}}
\put(257,48){\line(-3,-1){10}}
\put(242,43){\line(-3,-1){10}}
\put(227,38){\vector(-3,-1){10}}
\put(212,42){\tiny $\bar{\bf Q}_{-j}$}
\put(220,32){\line(3,1){10}}
\put(235,37){\line(3,1){10}}
\put(250,42){\line(3,1){10}}
\put(265,47){\vector(3,1){10}}
\put(270,40){\tiny ${\bf Q}^i_{-}$}

\put(272,113){\line(-3,-1){10}}
\put(257,108){\line(-3,-1){10}}
\put(242,103){\line(-3,-1){10}}
\put(227,98){\vector(-3,-1){10}}
\put(212,102){\tiny $\bar{\bf Q}_{-j}$}
\put(220,92){\line(3,1){10}}
\put(235,97){\line(3,1){10}}
\put(250,102){\line(3,1){10}}
\put(265,107){\vector(3,1){10}}
\put(270,100){\tiny ${\bf Q}^i_{-}$}

\put(272,173){\line(-3,-1){10}}
\put(257,168){\line(-3,-1){10}}
\put(242,163){\line(-3,-1){10}}
\put(227,158){\vector(-3,-1){10}}
\put(212,162){\tiny $\bar{\bf Q}_{-j}$}
\put(220,152){\line(3,1){10}}
\put(235,157){\line(3,1){10}}
\put(250,162){\line(3,1){10}}
\put(265,167){\vector(3,1){10}}
\put(270,160){\tiny ${\bf Q}^i_{-}$}

\put(5,26){$m$}
\put(0,56){$2m$}
\put(0,86){$3m$}
\put(0,116){$4m$}
\put(0,146){$5m$}
\put(0,176){$6m$}
\put(3,205){$\mbox{\textbf{\textit{H}}}$}

\put(55,0){$A_{+ i}$}
\put(110,0){$B$}
\put(180,0){$C_{(ik)}$}
\put(270,0){$A_{- i}$}

\end{picture}
\end{center}
\caption{The degeneracy of energy levels with respect to $\mbox{\textbf{\textit{H}}}={\bf H}-2m{\bf F}$ for $q\,{=}\,1/2$\,.
Circles and crosses denote bosonic and fermionic states, respectively. Solid lines correspond to the action of ${\bf Q}^i_{+}$ and $\bar{\bf Q}_{+j}$,
while dashed lines correspond to the action of ${\bf Q}^i_{-}$ and $\bar{\bf Q}_{-j}$.}
\label{figure2}
\end{figure}
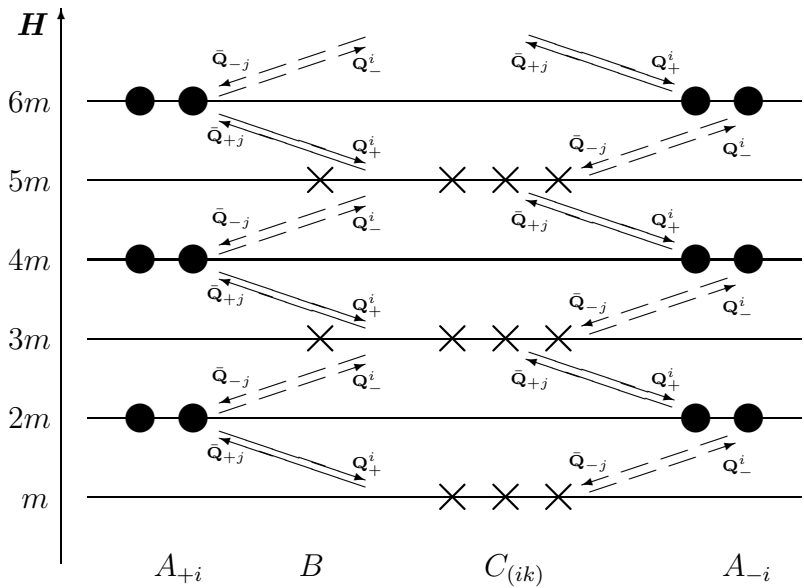

This mixing of states under the action of the supercharges $\mathbf{Q}^i_\pm$, $\bar{\mathbf{Q}}_\pm{}_i$
can be made more explicit and transparent, if we consider the eigenstates of the operator  $\mbox{\textbf{\textit{H}}}={\bf H}-2m{\bf F}$
[a quantum counterpart of (\ref{ca-H})]. Each energy level of the operator ${\bf H}$ (see Figure\,1)  is properly split
after passing to the energy levels of the operator $\mbox{\textbf{\textit{H}}}$: bosonic states $A_{\pm}$ acquire the additions $\pm m$ to their energy
and so give rise to additional (intermediate) levels in the spectrum of $\mbox{\textbf{\textit{H}}}$; the levels of the fermionic states remain unchanged.
The origin of such a splitting lies in the fact that the eigenstates of $\mbox{\textbf{\textit{H}}}$ do not
form irreducible ${\rm SU}(2|1)$ multiplets, because $\mbox{\textbf{\textit{H}}}$ does not commute even with the  ${\rm SU}(2|1)$ supercharges
$\mathbf{Q}^i_+$, $\bar{\mathbf{Q}}_+{}_i\,$.
The operators $\mathbf{Q}^i_\pm$, $\bar{\mathbf{Q}}_\pm{}_i$ mix adjacent bosonic and fermionic levels.
This picture for $q\,{=}\,1/2$ is given in Figure \ref{figure2}.

\section{Concluding remarks and outlook}

In this paper, we presented the quantum version  of the one-particle model of the multi-particle ${\rm SU}(2|1)$ supersymmetric mechanics
with additional semi-dynamical degrees of freedom which has been formulated in \cite{FI16}. Using the semi-dynamical  variables leads to
the considerable enrichment of the physical spectrum.
The relevant physical states are the ${\rm SU}(2)$ multi-spinors due to their dependence on spin variables.
We found that the states residing on a fixed energy level possess the fixed external ${\rm SU}(2)$ spin $q \in \big(\mathbb{Z}_{>0}$\,, $1/2 + \mathbb{Z}_{\geqslant 0}\big)$ and
form an irreducible ${\rm SU}(2|1)$ multiplet.
On the other hand, the states belonging to different energy levels with a fixed value of $q$
are naturally combined into an infinite-dimensional irreducible multiplet of
the more general ${\cal N}\,{=}\,4$ conformal supergroup  ${\rm OSp}(4|2)$. So this ${\rm OSp}(4|2)$ can be interpreted as a spectrum-generating supergroup for
our model.

In the future we plan to consider the quantum version  of a more general model of ${\rm SU}(2|1)$ supersymmetric mechanics defined by
the superfield action displaying an arbitrary dependence on dynamical {\bf (1,4,3)} superfield \cite{FI16}.
We hope that the corresponding quantum system with deformed supersymmetry will possess the hidden $D(2,1;\alpha)$ supersymmetry
for an arbitrary $\alpha$\,. Also, it is interesting to explore the quantum structure of the generic matrix model of ref. \cite{FI16},
which is a multi-particle generalization of the model considered here, and thus obtain a quantum realization of this novel ${\cal N}{=}\,4$
supersymmetric Calogero-Moser system.

\section*{Acknowledgements}

\noindent
This research was supported by the Russian Science Foundation Grant No 16-12-10306.


\end{document}